\documentclass[12pt]{JHEP3}
\usepackage{amsfonts,amssymb}
\usepackage{amsmath}

\usepackage{epsf}

\def\T11{{T}^{1,1}}
\def\bear{\begin{eqnarray}}
\def\eear{\end{eqnarray}}

\newcommand{\vac}{{|0\rangle}}

\newcommand{\pa}{\partial}
\newcommand{\tr}{{\rm tr}}
\newcommand{\comment}[1]{}

\newcommand{\pasl}{\pa\kern-.55em /}

\newcommand{\ksl}{k\kern-.55em /}

\newcommand{\ket}[1]{|#1\rangle}
\newcommand{\bra}[1]{\langle #1|}
\newcommand{\braket}[2]{\langle #1|#2\rangle}

\DeclareFixedFont{\xiiss}{OT1}{cmss}{m}{n}{12}
\DeclareFixedFont{\ixss}{OT1}{cmss}{m}{n}{9}
\DeclareFixedFont{\cmrnine}{OT1}{cmr}{m}{n}{9}
\newcommand{\field}[1]{\mathbb{#1}}

\newcommand{\BC}{{\field C}}
\newcommand{\BR}{{\field R}}
\newcommand{\BZ}{{\field Z}}

\newcommand{\CCs}{\hbox{\ixss C\kern-.4emI}}
\newcommand{\ZZs}{\hbox{\ixss Z\kern-.4emZ}}

\newcommand{\myfig}[3]{\begin{figure}[ht]
\begin{center}
\leavevmode \epsfxsize=#2cm \epsfbox{#1}
\end{center}
\caption{#3} \label{fig:#1}
\end{figure}}

\title{ Emergent geometry from $q$-deformations of ${\cal N}=4$
super Yang-Mills}
\author{David Berenstein$^{\dagger,1}$, Diego H. Correa$^{\ddagger,\sharp,2}$\\
$^\dagger$ Department of Physics, UCSB, Santa Barbara, CA 93106\\
$^\ddagger$ Kavli Institute for Theoretical Physics, UCSB, Santa
Barbara, CA 93106\\
 $^\sharp$ Centro de Estudios Cient\'{\i}ficos, Casilla
1469, Valdivia, Chile
\\
$^1$ \email{dberens@physics.ucsb.edu}
\\
$^2$ \email{dcorrea@cecs.cl}}

\abstract{We study BPS states in a marginal deformation of super
Yang-Mills on ${\mathbb R}\times S^3$ using a quantum mechanical
system of $q$-commuting matrices. We focus mainly on the case where
the parameter $q$ is a root of unity, so that the AdS dual of the
field theory can be associated to an orbifold of $AdS_5\times S^5$.
We show that in the large $N$ limit, BPS states are described by
density distributions of eigenvalues and we assign to these
distributions a geometrical spacetime interpretation. We go beyond
BPS configurations by turning on perturbative non-$q$-commuting
excitations. Considering states in an appropriate BMN limit, we use a saddle
point approximation to compute the BMN energy to all perturbative
orders in the 't Hooft coupling. We also examine some BMN like
states that correspond to twisted sector string states in the
orbifold and we show that our geometrical interpretation of the system
 is consistent with the quantum numbers of the corresponding states
under the quantum symmetry of the orbifold. }

 \preprint{NSF-KITP-05-91\\ CECS-PHY-05/13}

\begin{document}

\section{Introduction}

The AdS/CFT correspondence, in the most celebrated and understood
example, states the equivalence between the dynamics of type IIB
string theory on $AdS_5\times S^5$ and ${\cal N}=4$ super Yang-Mills
theory in four dimensions  \cite{Malda}.   According to the
correspondence, classical gravitational physics should be somehow
present in the strong 't Hooft coupling regime of the dual gauge
field theory. This means that one should be able to see the
emergence of  the $AdS_5\times S^5$ geometry and locality in the
strong 't Hooft coupling regime and large $N$ limit of ${\cal N}=4$
super Yang-Mills.

A first step towards the emergence of locality in the $S^5$ was
taken in \cite{toyads}, by studying the dynamics of 1/2 BPS
configurations of ${\cal N}=4$ SYM. This problem in the field theory
side is reduced to a matrix model quantum mechanics for a single
normal matrix and with first order dynamics. This system is
characterized by the matrix eigenvalues of this holomorphic matrix,
that turns out to be normal, and one is led to the study of $N$ free
two-dimensional fermions in a lowest Landau level system, perturbed
by a harmonic oscillator potential. Different 1/2 BPS configurations
are accounted for by different incompressible fermion droplets in a
two-dimensional phase space. The ground state of this system, {\it
i.e.} the circular droplet, is associated to the $AdS_5\times S^5$
background. In this picture, eigenvalues separated from the droplet
and holes in the droplet are supposed to represent D-branes in
$AdS_5\times S^5$. Separating several eigenvalues together from the
droplet corresponds to placing several D-branes on top of each
other. The stack of a large number of D-branes should be represented
by a supergravity solution.

 An impressive and surprising
confirmation of this picture was presented in \cite{LLM}, where all
regular 1/2 BPS solutions of type IIB supergravity are constructed
from some boundary conditions in a two-dimensional section that
coincides with the picture of fermion droplets in a two-dimensional
space. We mentioned that one can start to see the emergence of
locality in the $S^5$ in this picture. In this regard, the edge of
the circular droplet is identified with an equatorial circle of the
$S^5$, and for more general solutions, the plane that describes the
fermion droplet configurations is a submanifold of the supergravity
solution.

Similar ideas were put forward in \cite{Droplet}, but in this case
1/4 and 1/8 BPS configurations of ${\cal N}=4$ SYM were considered
instead.
 For 1/8 BPS configurations, the field theory problem is reduced
to a matrix model quantum mechanics for three commuting normal
matrices. This model is arrived at by realizing that the BPS
configurations of the field theory on $S^3$ are related to the
moduli space of vacua of the field theory. For ${\cal N}=4$ SYM this
means that one has to consider systems that satisfy the $F$-term and
$D$-term constraints, and this leads us to the study of six
hermitian commuting matrices, that are paired into three normal
matrices once a choice of a particular ${\cal N}=1$ supersymmetry has been
made.

Again the dynamics is described in terms of the eigenvalues of
matrices, but they cannot be accounted for as free fermions anymore.
Instead one ends up describing the system by holomorphic wave functions of $N$
six-dimensional bosons in a harmonic oscillator potential, subject
to a repulsive interaction. This effective repulsion arises from
measure terms in the transformation from general commuting matrices
to diagonal matrices and it is the usual repulsion of eigenvalues
that arises from the Vandermonde determinant in matrix models. In
the large $N$ limit, one can think of the system as a thermodynamic
system and obtain a density distribution of eigenvalues in a
six-dimensional phase space. For the ground state, the eigenvalues
are uniformly distributed in a round hollow $S^5$. This was argued
to correspond exactly to the $S^5$ near horizon factor in the dual
geometry.

If non-BPS modes of finite energy are turned on, they are associated
to off-diagonal modes of the matrices. These off-diagonal modes to a
first approximation can be treated perturbatively  and they become
bi-local in this $S^5$ geometry.  This is because they can be drawn
as an arrow connecting two different eigenvalues, that become
associated geometrically with points on the sphere (the sphere
depicts the distribution of eigenvalues after all). In this setup
they can be interpreted as bits of massive strings and the $S^5$ can
be directly identified with the $S^5$  of the geometry $AdS_5\times
S^5$ that sits at the center of AdS space in global coordinates.
This picture can be used to successfully reproduce geometrical
calculations for some moving strings. For instance, the energy of
BMN string states \cite{BMN} can be obtained in this picture by a
simple saddle point approximation \cite{BCV}. This formalism for
doing calculations presents a new viewpoint of how the AdS spacetime
geometry is encoded in the dual CFT for the particular case of $N=4
$ SYM on $S^3\times \BR$.

It is natural to ask if this description of the emergence of
geometry and gravity from SYM is unique to ${\cal N}=4 $ SYM or not.
In this paper we show that the emergence of gravity from the matrix
model describing the BPS configurations of the field theory can be
extended to other realizations of the AdS/CFT correspondence. What
changes with respect to ${\cal N}= 4$ is that the set of matrices
that we need to consider do not commute any longer. However, they do
not differ significantly from commuting matrices, so that one can
still talk about distributions of eigenvalues that are given by
block diagonal sets of matrices describing the scalar fields of the
theory. This is related to the notions of geometry arising from
describing moduli spaces of D-branes at Calabi-Yau singularities, as
explained in \cite{Rev}.

Clearly, the simplest examples of such geometries are orbifolds of
${\cal N}=4 $ SYM \cite{KacS}, and one would want to study the
dynamics of branes at these singularities with the techniques
developed in \cite{Droplet}.

Some of these orbifolds give very simple theories that have the same
field content as ${\cal N}=4 $ SYM and are natural candidates to
investigate. These turn out to be field theories on the moduli space
of marginal deformations of ${\cal N}=4$ SYM, preserving ${\cal
N}=1$ supersymmetry \cite{LS}. These particular deformations of
${\cal N}=4 $ SYM have a surprisingly rich dynamics and have been
extensively studied in the past. The gravity duals for those
deformations with $U(1)^3$ global symmetry were recently constructed
\cite{LM}, and this has sparked new interest in the study of these
field theories. Also, when the deformation parameter $q$ takes
special values, the gravity dual is simply the near horizon of
D-branes in orbifolds with discrete torsion \cite{Douglas, DF,BL},
which is simply $AdS_5\times S^5/\Gamma$, where $\Gamma$ is a
particular discrete symmetry group $\BZ_n\times \BZ_n$ for $q$ given
as an $n$-th root of unity.

We will consider these deformations in the particular case where $q$
is a root of unity and we will show the emergence of the orbifolded
geometry from a matrix model, in a similar vein to the derivation of
the geometry of the five sphere in the $AdS_5\times S^5$ case. We
also obtain within this picture the all loop anomalous dimension of
the BMN limit of states in the Lunin-Maldacena background and the
orbifold geometries, plus we show that some peculiarities of the
saddle point evaluation of energies of these states are required for
consistency of the geometric interpretation as string states in
$AdS_5\times S^5/\Gamma$ with some fixed quantum numbers with
respect to the quantum symmetry of the orbifold.

The paper is organized as follows. In section \ref{mqm} we derive a
matrix model of $q$-commuting matrices describing the BPS
configurations of a special deformation  of ${\cal  N}=4$ SYM  and
show the emergence of the orbifolded $S^5$ as a density distribution
of eigenvalues. In section \ref{bits} we show that non-BPS modes in
this setup are localized in the orbifolded $S^5$ and interpreted as
massive string bits. In section \ref{spa}  we compute the energy of
BMN states to all perturbative orders, and find that
similar states that are in the twisted sector correspond to certain
non-trivial loops in the orbifold geometry that lift to open paths
in the covering space. In section \ref{dis} we discuss and summarize
the results of this paper.

\section{Quantum mechanics for $q$-commuting matrices}
\label{mqm}

To begin, we will study a one parameter set of  marginal
deformations of ${\cal N}= 4$ SYM that has $U(1)^3$ global symmetry.
These were originally shown to be conformal by Leigh and Strassler
\cite{LS}. These depend on a complex parameter $q$, that we will
further specialize to be a root of unity, and they present the
following superpotential for the three adjoint superfields $X$, $Y$
and $Z$,
\begin{equation}
W_q = {\rm tr}(XYZ)-q{\rm tr}(XZY)\,.
\end{equation}
When the deformation parameter $q=r\exp(2\pi i \beta)$ takes the
special values $r=1$ and $\beta$ rational, the CFT is related to
geometries of orbifolds with discrete torsion \cite{Douglas, DF,BL}.
For simplicity we will focus exactly on these deformations such that
$q$ is a primitive $n$-th root of unity, and we will try to
 see how our calculations depend on $q$. Our goal is to be able to say
that the ground state of the  CFT dual of these theories is given
geometrically by $AdS_5\times S^5/\Gamma$ and that we can reproduce
part of the spectrum of strings on this geometry from our
considerations. Our ultimate goal is to show that the techniques
developed in \cite{Droplet,BCV} can be applied in a broad class of
examples and are not restricted to having a quantum system with 32
supersymmetries.

The line of thought explored in \cite{Droplet} begins by considering
the chiral ring of the field theory on Euclidean space, and trying
to understand what states are associated to the operators in the
chiral ring under the operator state correspondence. Chiral ring
operators preserve some of the supersymmetries, so the dual states
are BPS states. Therefore,  in the AdS/CFT correspondence we want to
consider BPS configurations of the field theory on $S^3$, that are
dual to the set of chiral primary operators. These operators, in the
free field limit can only be built from the constant modes of the
fields on the $S^3$. One can additionally make some other such
states by turning on some of the fermion fields. However these are
zero in a naive semiclassical treatment, so we will ignore them for
the purposes of this paper. From these considerations we get an
effective dimensional reduction to a matrix model whose degrees of
freedom are the constant scalar modes of the fields on the $S^3$.
Focusing on these modes alone, we get the following effective action
\begin{equation}
S = \pi^2\!\int dt \,  {\rm tr} \left(|D_t X|^2 + |D_t Y|^2+ |D_t
Z|^2-|X|^2-|Y|^2-|Z|^2 - V_D -  V_F\right)\, ,
\end{equation}
where $V_D$ and $V_F$ are the potentials resulting from integrating
out the auxiliary D-terms and F-terms in the ${\cal N}=1$
lagrangian. The factors of $\pi$ result from integration over the
volume of the $S^3$. Moreover, the BPS constraint is of the form
\begin{equation}
\Delta = J
\end{equation}
where $\Delta$ is the Hamiltonian of the field theory (and also the
generator of dilatations of the Euclidean field theory under the
operator-state correspondence map), while $J$ is the generator of
the R charge with a slightly different normalization that matches
the BMN conventions \cite{BMN}. In the free field limit we satisfy
this constraint by taking positive frequency solutions of the
holomorphic fields of the ${\cal N}=1$ theory and in this limit we
ignore $V_D$ and $V_F$ terms in the action. With the normalizations
above we get that the solutions have the following time dependence
\begin{eqnarray}
X(t)&=& X(0)\exp(i t)\\
Y(t)&=& Y(0) \exp(i t)\\
Z(t)&=& Z(0)\exp(i t)
\end{eqnarray}

To consider classical solutions of the BPS constraints in the
interacting case, we notice that the solutions above satisfy
$\Delta=J$ if one also imposes the constraints $V_D=V_F=0$. In the
usual ${\cal N}=1$ supersymmetric field theories, this corresponds
to the classical moduli space problem of finding the supersymmetric
vacua of the supersymmetric field theory. For theories associated to
D-brane actions, this can be solved in two steps. First solve the
$F$-terms, that are given by matrix equations, and afterwards on the
set of these solutions one does complexified gauge transformations
to try solve the $D$-term constraints.

For this particular field theory, the characterization of the moduli
space of supersymmetric vacua of these marginal deformations  have
been studied in \cite{BL,BJL}. The classical moduli space is
determined by solutions of the $F$-terms constraints, up to
$GL(N,\BC)$ equivalence (the complexification of the gauge group, as
is familiar in supersymmetric theories)
\begin{eqnarray}
XY -q YX&=&0\, , \nonumber\\
YZ -q ZY&=&0\, , \nonumber\\
ZX -q XZ&=&0\, . \label{qal}
\end{eqnarray}
This means that we have to look for $N\times N$ $q$-commuting
matrices. We are considering the special case $q=\exp(2\pi i/n)$.
For this case, an $n$-dimensional representation of the algebra
(\ref{qal}) can be obtained in terms of the clock and shift matrices
$P$ and $Q$
%
\begin{equation}
P=\left(\begin{array}{cccc}
1&0&\ldots & 0 \\
0&q&\ldots & 0 \\
\vdots&\vdots&\ddots&\vdots \\
0&0&\ldots & q^{n-1} \\
\end{array}\right)
\, , ~ ~ ~ Q=\left(\begin{array}{cccc}
0&0&\ldots & 1 \\
1&0&\ldots & 0 \\
\vdots&\ddots&\ddots&\vdots \\
0&0&1 & 0 \\
\end{array}\right)\, .
\end{equation}
The algebra (\ref{qal}) is solved by the following representation:
simply take $X\sim P$, $Y\sim Q$ and $Z\sim P^{n-1}Q^{n-1}$, and
this is an $n$-dimensional irreducible representation of the
algebra. Noticing that $X^n, Y^n, Z^n, XYZ$ are proportional to the
identity on these representations, one can identify these as
coordinates of a commutative geometry, that turns out to be exactly
the ring of holomorphic functions on the orbifold $\BC^3/\BZ_n\times
\BZ_n$.

There are also branches of fractional branes where two out of the
three matrices are zero $X=Y=0$ lets say \cite{BL,BJL}. These can
have more moduli associated to them as the eigenvalues of $Z$ become
unrelated to each other \footnote{ These branches have been studied
non-perturbatively in the work of Dorey and collaborators, see
\cite{DHK,Dorey:2003pp,BDor} and references therein.}.

However,  it is easy to show that their orbits in the configuration
space of matrices (ignoring the equivalence of representations under
$GL(N,\BC)$) always have higher codimension than the matrices shown
above. Since we are trying to understand the ground state of the
matrix model we discussed, one can argue that these branches are
less important because they have higher codimension on the set of
allowed matrix configurations. We will ignore these branches in what
follows, assuming that for the ground state wave function, the other
branch made only of branes in the bulk dominates. This will be shown
to be self-consistent later on in the paper. Moreover, these extra
fractional brane branches are only present for singular three
dimensional CY spaces where the singularity at which the branes are
placed is not isolated, so they are not relevant for isolated
singularities.

Let us now also assume for simplicity that the rank of the gauge
group is $N= m n$. In this case, it is straightforward  to construct
an $N$-dimensional representation of the same algebra, with tensor
products of matrices \cite{DHK},
\begin{eqnarray}
X&=& \frac{1}{\sqrt{2\pi^2 n}}\ \Lambda^1\otimes P\, , \label{qal1}\\
Y &=&\frac{1}{\sqrt{2\pi^2 n}} \ \Lambda^2\otimes Q\, ,\label{qal2}\\
Z &=&\frac{1}{\sqrt{2\pi^2 n}}\ \Lambda^3\otimes P^{n-1}Q^{n-1}\, ,
\label{qal3}
\end{eqnarray}
where $\Lambda^1$, $\Lambda^2$ and $\Lambda^3$ are three commuting
holomorphic $m\times m$ matrices. These solutions are just direct
sums of the irreducible representations found above, and they solve
the $F$-term equations automatically. Indeed, one can show that the
generic representation of $m$ D-branes in the bulk away from the
singularities of the geometry corresponds exactly with such a
representation, and that any other general solution
 can be transformed into this form by a gauge transformation. This is explained in detail in \cite{BL,Rev}. Here
we will just use these results. In \cite{BL,Rev} was also noted that there are special solutions
of representations where two of $X,Y,Z$ are zero and the third is a c-number. 
These correspond to fractional branes at the singular locus. These solutions are 
ignored above.

The normalization factor above for the parametrization of the
matrices was included for later convenience (this normalization
guarantees that the action for the matrices $\Lambda^i$ has a
canonical normalization when the theory is compactified on the
sphere).

Now, the moduli space of supersymmetric vacua is parametrized by $3
m$ eigenvalues of the commuting matrices $\Lambda^1$, $\Lambda^2$
and $\Lambda^3$. We should notice that, if we call this
representation $R(\Lambda^1,\Lambda^2,\Lambda^3)$, representations
$R(q\Lambda^1,q^{-1}\Lambda^2,\Lambda^3)$ and
$R(q\Lambda^1,\Lambda^2,q^{-1}\Lambda^3)$ are equivalent under
similarity transformations. This is, these configurations are
equivalent under gauge transformations. Imposing this equivalence,
the eigenvalues parametrizing the moduli space take values in the
orbifold ${\mathbb C}^3/{\mathbb Z}_n\times{\mathbb Z}_n$. This
procedure can be implemented at the level of wave functions for the
matrices by requiring that the wave function is invariant under
these identifications of configurations.

We can now obtain a matrix model from the scalar sector of
the ${\cal N}=1$ marginal deformation of ${\cal N}=4$ SYM action in
question. We define the field theory in ${\mathbb R}\times S^3$ and
study BPS configurations keeping  only the $s$-waves modes of the
$q$-commuting scalar fields (\ref{qal2}). This follows the
techniques used in \cite{Droplet}, where these classical states
satisfying the $F$-term constraints on the $S^3$ are related to
operators on the chiral ring of the field theory.

To obtain the matrix model we write an ansatz with spherical
symmetry for the fields that respect these $F$-term constraints into
the effective action of the field theory on $S^3$.  For solutions
with this ansatz, that moreover satisfy the $D$-term constraints as
well,  the $D$-term and $F$-term potential term in the SYM action on
$S^3$ vanishes. The solutions as written above satisfy all the
necessary constraints, provided that the $\Lambda^i$ matrices are
normal matrices and that they commute with each other. These fields
are still massive in the reduced matrix model because these scalars
are conformally coupled to the round metric on the $S^3$. We then
perform the integral over the $S^3$ and we get a quantum mechanics
matrix model with a quadratic potential,
\begin{eqnarray}
S &=& 2\pi^2\int\, dt \, \frac 12 {\rm tr}
\left(|D_t X|^2 + |D_t Y|^2+ |D_t Z|^2-|X|^2-|Y|^2-|Z|^2\right)\nonumber\\
&=& \int\, dt \, \frac 12 {\rm tr} \left(|D_t \Lambda^1|^2 + |D_t
\Lambda^2|^2+ |D_t
\Lambda^3|^2-|\Lambda^1|^2-|\Lambda^2|^2-|\Lambda^3|^2\right)\, .
\end{eqnarray}
The normalization of the matrices cancels factors of $n$ from the
trace of the identity on the set of $P,Q$ matrices. Since the
$\Lambda^i$ matrices are mutually commuting, and normal, we can
diagonalize them simultaneously with a unitary transformation,
together with ${\bar \Lambda}^i$.

This is a gauge transformation on the set of configurations. This
reduces the degrees of freedom from generic commuting matrices to
their eigenvalues. Classically, the system is reduced to $m$
decoupled harmonic oscillators in three dimensions.

However, the emergence of gravity as detailed in \cite{Droplet}
takes place in the quantum mechanical matrix model. The reduction of
degrees of freedom to the eigenvalues of the $\Lambda^i$ matrices is
valid only if we include the measure factors coming from the volume
of the gauge orbit. Because of this measure, the harmonic
oscillators will no longer be decoupled. Since there are global
transformations that permute the eigenvalues, the wave functions
also have to be symmetric under these permutations. As done in
\cite{Droplet}, this system will be interpreted as a set of $m$
interacting bosons in a 6 dimensional phase space. This requires
looking at the BPS constraint, that only keeps the positive
frequency modes of the $\Lambda^i$ matrices. This procedure
effectively converts the classical moduli space into a symplectic
geometry, because for these solutions $\pi_{X,Y,Z}\sim {\bar X, \bar
Y, \bar Z}$ and one has an effective magnetic field on the moduli
space that makes the problem similar to a quantum hall system: we
want to quantize only the lowest Landau level problem.

Describing the $q$-commuting matrices as (\ref{qal2}) with diagonal
$\Lambda^i$ is a gauge choice. The quantum modifications of the free
Hamiltonian that we need arise from computing measure effects from
changes of variables. In the original variables, the Laplacian of the
Hamiltonian has a cartesian coordinate form. Taking into consideration
measure effects is very similar to the problem of writing the Laplacian of
Euclidean flat space in spherical coordinates, using the integration
measure of spherical coordinates.

The measure terms we are after come from calculating the volume of
the gauge orbits in question. To compute this volume we set matrices
in the form (\ref{qal2}) with diagonal $\Lambda^i$ and perform an
infinitesimal gauge transformation with the broken generators, {\it
i.e.} transformations that turn on off-diagonal components in
matrices $\Lambda$ and give rise to general combinations $P^\alpha
Q^\beta$ in the tensor products \footnote{To refer to different
kinds of indices we use different sets of letters. Indices
$a,b,\ldots = 1,\ldots,m$ and indices $\alpha,\beta,\ldots
=1,\ldots, n$.}. These variations are
\begin{eqnarray}
\delta X &=&
q^{-\beta}\theta^{\alpha\beta}_{ab}(\lambda_b^1-q^\beta\lambda_a^1)
\Gamma_{ab}\otimes P^{\alpha+1}Q^\beta\, , \nonumber\\
\delta Y
&=&\theta^{\alpha\beta}_{ab}(\lambda_b^2-q^{-\alpha}\lambda_a^2)
\Gamma_{ab}\otimes P^{\alpha}Q^{\beta+1}\, , \nonumber\\
\delta Z &=&q^\beta
\theta^{\alpha\beta}_{ab}(\lambda_b^3-q^{\alpha-\beta}\lambda_a^3)
\Gamma_{ab}\otimes P^{\alpha-1}Q^{\beta-1}\, , \label{vari}
\end{eqnarray}
where $\theta^{\alpha\beta}_{ab}$ parametrize the transformations
generated by $\Gamma_{ab}\otimes P^{\alpha}Q^{\beta}$ and
$\Gamma_{ab}$ is a $m\times m$ matrix whose components are
$(\Gamma_{ab})_{cd}=\delta_{ac}\delta_{bd}$.

As we said, the vectors of complex eigenvalues $\vec\lambda_a =
 (\lambda^1_a,\lambda^2_a,\lambda^3_a)$  take values in the complex orbifold
${\mathbb C}^3/{\mathbb Z}_n\times{\mathbb Z}_n$ because of gauge
identifications between different configurations. Roughly, we could
plot all these eigenvalues in a $n^{2}$-th wedge of the whole
${\mathbb C}^3$. For expressing the measure, it is convenient to
distinguish among the $n^2$ images in the complete ${\mathbb C}^3$.
So, we define $\vec\lambda_a^{\alpha,\beta}=(\lambda^1_a q^{\alpha
+\beta}, \lambda^2_aq^{-\alpha},\lambda^3_a q^{-\beta})$. The volume
associated to the variation $\theta^{\alpha\beta}_{ab}$ is equal to
the length of the vector $(\vec \lambda_a-\vec
\lambda_b)^{\alpha,\beta}$. Then, the measure turns out to be
\begin{equation}
\mu^2 = {\prod_{\alpha,\beta}}'\prod_{a}|\vec \lambda_a-\vec
\lambda_a^{\alpha,\beta}| \cdot \prod_{\alpha,\beta}\prod_{a<
b}|\vec \lambda_a-\vec \lambda_b^{\alpha,\beta}|^2 \; . \label{me}
\end{equation}
The prime in the first product is to remark that the case
$\alpha=\beta=n$ is excluded. At this point, we can define new
indices $A,B,\ldots$ running from 1 to $m n^2$. We also call
$\vec\lambda_A$ the $m n^2$ vectors $\vec \lambda_a^{\alpha,\beta}$.
Moreover, the measure (\ref{me}) resembles the vectorial
generalization of the Vandermonde determinant that appeared in
\cite{Droplet}, evaluated for the $mn^2$ vectors $\vec\lambda_A$.
However, many $|\vec\lambda_A-\vec\lambda_B|^2$ are missing. In the
complete Vandermonde determinant factors are repeated because only
$m$ of the $mn^2$ vectors  $\vec \lambda_a^{\alpha,\beta}$ are
arbitrary, while the rest are the images under the action of
${\mathbb Z}_n\times{\mathbb Z}_n$. Measure (\ref{me}) does not
present such a repetition of factors and coincides exactly with the
$n^2$-th root of this determinant, where each different length is
counted only once
\begin{equation}
\mu^2  = \left(\prod_{A<B}\!|\vec \lambda_A-\vec
\lambda_B|^2\right)^{1/n^2} \; . \label{me2}
\end{equation}
We could have guessed this form of the measure by using the method
of images on the covering space and taking the $n^2$-th root of the
result to avoid over-counting of states.

Now, we use this measure factor to define the reduced Hamiltonian in
the eigenvalue basis
\begin{equation}
H= \sum_a -\frac 1{2\mu^2} \nabla_a \mu^2 \nabla_a + \frac 12 |\vec
\lambda_a|^2 \;. \label{hami}
\end{equation}
We will look for the ground state wave function of Hamiltonian
(\ref{hami}) and give a probabilistic interpretation of its square
modulus as a Boltzman distribution in the limit of large $m$. As in
the undeformed case \cite{Droplet}, we can show that the ground
state for $m$ decoupled  harmonic oscillators,
\begin{equation}
\psi_0 \sim \exp(-\sum |\vec\lambda_a|^2/2)\;, \label{ge}
\end{equation}
is an exact eigenfunction of (\ref{hami}) that has the correct
symmetry under exchange of particles and also under the discrete
group orbit $\Gamma$. In what follow, and to simplify the notation, we write the vectors
of 3 complex dimensions $\vec\lambda_a$ as vectors of 6 real
dimensions $\vec x_a$. Moreover, the vector gradient
$\vec\nabla_a$ now refers to the gradient associated to the 6 real
coordinates. Using that $\vec\nabla_a\psi_0=-\vec x_a\psi_0$, we have
\begin{eqnarray}
H\psi_0 &=& \sum_a \frac 1{2\mu^2} \vec\nabla_a (\mu^2 \vec
x_a\psi_0)
 + \frac 12 |\vec x_a|^2\psi_0 \nonumber\\
&=& \sum_a \frac 1{2\mu^2} (\vec x_a \vec\nabla_a \mu^2)\psi_0
 + 6 m \psi_0 \;. \label{hami2}
\end{eqnarray}
As in the undeformed case, following \cite{Droplet}, the measure is
a homogeneous function, now of degree $m(mn^2-1)$. Then, $\mu^2$ is
an eigenfunction of the operator $\sum\vec x_a \vec\nabla_a$ and
$\psi_0$ is an eigenfunction of the reduced Hamiltonian. Since,
$\psi_0$ is real and positive one expects this eigenfunction  to be
the ground state of the system. The states dual to primary fields
should arise from multiplying the above ground state by holomorphic
polynomials in the $X,Y,Z$ fields, subject to the moduli space
constraints. This effectively results in a subset of the polynomials
of the $\Lambda^i$, because the traces also include traces over the
$P,Q$ matrices part. These vanish unless the corresponding trace is
proportional to the identity.

Orthogonality relations with another eigenfunction and expectation
values of observables have to be computed using the integration
measure $\mu^2$ we have found. For that reason, it is convenient to
absorb a $\mu$ factor into the definition of the wave function,
\begin{equation}
\hat \psi = \mu \psi\;.
\end{equation}
Then, after this similarity transformation, we get the usual
integration measure $\prod d^{6}x_a$ associated to wave functions
$\hat\psi$. The factor $\mu$ is also symmetric under the exchange of
all the vectors $\vec x_a$ and under the discrete identifications of
eigenvalues. So the particles are also $m$ identical bosons with
respect to the new measure.  To consider the emergence of geometry,
we are instructed to place the $m$ bosons in the same phase space
and study the density distributions of particles for the given
wave-functions described by the $\hat \psi$. We can consider the
modulus square of a given wave function as a probability
distribution of the $m$ bosons in the 6 dimensional space ${\mathbb
C}^3/{\mathbb Z}_n\times{\mathbb Z}_n$. For the wave function
 $\psi_0$ we have
\begin{equation}
|\hat \psi_0^2| \sim \exp \left(-\sum_a \left(|\vec x_a|^2 -
 {\sum_{\alpha,\beta}}'\log|\vec x_a-\vec x_a^{\alpha,\beta}|\right)+
2\sum_{a< b}\sum_{\alpha,\beta}\log|\vec x_a-\vec
x_b^{\alpha,\beta}| \right)\;.
\end{equation}
The computation of position observables from the square of the wave
function is equivalent to a statistical mechanics problem of
calculating the partition function for $m$ bosons in an external
quadratic potential. These bosons are subject to a logarithmic
repulsion between them and with respect to their images through the
action of ${\mathbb Z}_n\times{\mathbb Z}_n$. We consider the
thermodynamic limit  $m \to \infty$, where the bosons form some
continuous and positive distribution density $\rho$ on the phase
space of a single particle and look for the most probable $\rho$. In
this continuous limit, sums over $a$ are converted into integrals
over ${\mathbb C}^3/{\mathbb Z}_n\times{\mathbb Z}_n$, or integrals
over $\BC^3$ if we take care of normalizing the density at a point
and its images properly.
\begin{eqnarray}
\sum_a (\ldots ) ~ & \to& ~ \int_{{\mathbb C}^3/{\mathbb
Z}_n\times{\mathbb Z}_n} \!\!\!\!\!\!\!\!\!\!\!\!\!\!\!\! d^6x\
\rho(x) (\ldots )
\\m &=& \int_{{\mathbb C}^3/{\mathbb Z}_n\times{\mathbb Z}_n}
\!\!\!\!\!\!\!\!\!\!\!\!\!\!\!\! d^6x\  \rho(x) \label{cons}
\end{eqnarray}
The probability distribution in terms of the density $\rho$ is
\begin{eqnarray}
|\hat \psi_0^2| &\sim & \exp \left( -  \int_{{\mathbb C}^3/{\mathbb
Z}_n\times{\mathbb Z}_n} \!\!\!\!\!\!\!\!\!\!\!\!\!\!\!\!\!\!
d^6x\rho(x) \vec x^2 + \int_{{\mathbb C}^3/{\mathbb
Z}_n\times{\mathbb Z}_n} \!\!\!\!\!\!\!\!\!\!\!\!\!\!\!\!\!\! d^6x
d^6y \rho(x)\rho(y) {\sum_{\alpha,\beta}}\log|\vec x-\vec
y^{\alpha,\beta}|\right) \;.\label{eq:var}
\end{eqnarray}
Maximizing the probability density (\ref{eq:var}) we can find the
most likely boson distribution, exactly as in \cite{Droplet}.
Varying the argument of the exponential we obtain the following
integral equation for $\rho(x)$
\begin{equation}
\vec x^2 + C = 2 \int_{{\mathbb C}^3/{\mathbb Z}_n\times{\mathbb
Z}_n} \!\!\!\!\!\!\!\!\!\!\!\!\!\!\!\!\!\! d^6y\ \rho_0(y)
{\sum_{\alpha,\beta}}\log|\vec x_a-\vec y_a^{\alpha,\beta}|\;.
\label{eom}
\end{equation}
The constant $C$ is a Lagrange multiplier enforcing the constraint
in the total number of bosons (\ref{cons}). In 6 dimensions the
function $\log| \vec x - \vec y|$ is proportional to the Green's
function for the operator $(\nabla^2)^3$. If we operate naively on
both sides of (\ref{eom}) with this operator we obtain  $\rho_0 =0$
. From this expression we conclude that the distribution $\rho_0$
has singular support, because the naive manipulations are only valid
in the case that $\rho$ is differentiable. In the covering space of
the orbifold, we see a spherically symmetric Boltzman gas of
particles, identical in nature to that one found in \cite{Droplet}.

Because of symmetry, the simplest ansatz for $\rho_0$ we can make is
a singular spherically symmetric distribution at radius $r_0$ on
${\mathbb C}^3/{\mathbb Z}_n\times{\mathbb Z}_n$. So, the
eigenvalues of the ground state are uniformly distributed in an
orbifolded 5-sphere $S^{5}/{\mathbb Z}_n\times{\mathbb Z}_n$,
\begin{equation}
\rho_0 = m \frac{\delta(|\vec x|- r_0)} {r_0^{5} {\rm Vol}
(S^{5}/{\mathbb Z}_n\times{\mathbb Z}_n)}\;, \label{rho0}
\end{equation}
where the normalization was chosen so that the constraint
(\ref{cons}) is fulfilled.  It is expected that this is the unique saddle point for the 
distribution of the density of eigenvalues (this has not been proved yet for this problem).

This $S^{5}/{\mathbb Z}_n\times{\mathbb Z}_n$ can be identified with
the compact factor of $AdS_5\times S^{5}/{\mathbb Z}_n\times{\mathbb
Z}_n$, which is the geometrical background of the string theory dual
to the $q$-deformation of SYM we are considering \cite{BL,BJL}. In
the next section, we will show how the energy of simple non-BPS
excitations are localized in the $S^{5}/{\mathbb Z}_n\times{\mathbb
Z}_n$ and we will be able to interpret them in terms of bits of
massive strings.

Now we substitute this ansatz into the argument of the exponential
(we call it ${\cal H}$) in (\ref{eq:var}), and minimize with respect
to $r_0$. Since the distribution is supported at $r_0$ and uniformly
distributed in the angular variables, it is easy to isolate the
$r_0$ dependence. We obtain,
\begin{equation}
{\cal H}=mr_0^2 - m^2 n^2\log r_0 +\kappa\, ,
\end{equation}
where $\kappa$ is a very complicated integral on the angles of the
$S^{5}/{\mathbb Z}_n\times {\mathbb Z}_n$ but independent of $r_0$.
Now, minimizing ${\cal H}$ with respect to $r_0$ we easily find the
critical radius
\begin{equation}
r_0 = \sqrt {\frac {mn^2}2}=\sqrt {\frac {n N}2}\;. \label{rad}
\end{equation}
This result is relevant for the computation of BMN excitation
energies  we will carry through in section \ref{spa}. At first sight
it seems that this result depends on $n$. However, this is an
artifact of the normalization factor of $n$ used in (\ref{qal3}). If
we restore the standard matrix normalization, we find that the
radius of this sphere is independent of $n$ (this translates into
the radius being independent of $q$, for $q$ unitary). However, some
of the intermediate steps to derive the result would involve factors
of $n$ in various other places (the Hamiltonian and the wave
function for example). This result is crucial for us. This is what
ensures that calculations can be analytic functions of $q$. For
analytic functions, knowing the rational values of the function lets
us find the result for arbitrary values of the phase of $q$ by
continuity. This is what will  let us extend some of our analysis to
the non-BPS case and to explore states in the Lunin-Maldacena
background \cite{LM}.

Most of the other BPS wave functions of the system are given by
holomorphic functions of the $\Lambda$ multiplying $\hat\psi_0$ that
satisfy the boson statistics and that are invariant under the
identifications produced by $\Gamma$. These symmetries are all
required from the residual gauge identification of configurations.
In general, one obtains in this way the chiral ring associated to
gravity modes in the bulk of the spacetime (untwisted states in the
chiral ring). These can also be interpreted as wave functions in the
covering space. If one follows the gravity interpretation of these
states in the $AdS_5\times S^5$ case, one finds that coherent states
of single trace operators correspond to deformations of the $S^5$
geometry that respect the discrete symmetries of the orbifold, and
hence they correspond to supergravity solutions that deform the
shape of the $S^5/\Gamma$.

To include also the twisted sector BPS states, one also needs to
consider having some small fraction of the eigenvalues to be
located in the fractional brane branches. Understanding exactly how
this works is beyond the scope of the present paper, but it is a
very interesting problem to consider, as these deformations smooth
out the singularities of the $S^5/\Gamma$ space.
 Some progress for solutions with branes on the Coulomb branch
 of the field theory has been made
at the level of supergravity solutions with Lunin-Maldacena boundary
conditions in \cite{AV,HSZ}, but these preserve many symmetries that
make the configurations non-generic.

\section{String bits}
\label{bits}

In this section, we want to describe the emergence of massive
stringy modes in the CFT as in \cite{BCV}, see also \cite{Rod} for
an alternative viewpoint on some of these computations. The first
thing to do is to go beyond BPS configurations. So, let us include
in the action the potential $D$-terms and $F$-terms and consider
situations where we are in a near BPS state. The action for the
$s$-wave scalar modes is

\begin{eqnarray}
S = \pi^2\!\int dt \,  {\rm tr} \left(|D_t X|^2 + |D_t Y|^2+ |D_t
Z|^2-|X|^2-|Y|^2-|Z|^2 - V_D -  V_F\right)\, ,
\end{eqnarray}
where
\begin{eqnarray}
V_D &=&\frac{g_{YM}^2}{8}{\rm tr}\left(|[X,\bar X]+[Y,\bar
Y]+[Z,\bar Z]|^2\right)\,,
\\
V_F &=&\frac{g_{YM}^2}{2}{\rm tr}\left(|[X,Y]_q|^2
+|[Y,Z]_q|^2+|[Z,X]_q|^2\right)\,.
\end{eqnarray}
Being near BPS, means that we can attempt to treat the degrees of
freedom that take us away from being BPS as a perturbation. These
should be given by infinitesimal deformations of BPS configurations.
These configurations result from turning on the other matrix
components of the fields $X,Y,Z$ that can not be turned on in a BPS
configuration. In perturbation theory we treat these non-BPS modes
as fast degrees of freedom attached to pairs of eigenvalues. For a
non-BPS component in the matrix position $a,b$, one can associate the
infinitesimal matrix component to the eigenvalues of the matrix in
the diagonal components $a$ and $b$. The eigenvalues in this setup
are treated as slow degrees of freedom. As a first approximation, we
treat the off-diagonal modes as free fields coupled to a background
of eigenvalues, and we ignore the back-reaction of the eigenvalues
to the string bits.

In order turn on a non-BPS mode we should include in at least one of
the holomorphic fields $X,Y,Z$ an off-diagonal component in the
matrices $\Lambda^1,\Lambda^2,\Lambda^3$ and/or a general $P^\alpha
Q^\beta$ matrix in the second factor of the tensor product. Let us
consider, for instance,
\begin{equation}
X'=X+\delta X^{(\alpha,\beta)}_{(a,b)}
\end{equation}
where $X$ is the same as in (\ref{qal2}) and the non-BPS piece is
\begin{equation}
\delta X^{(\alpha,\beta)}_{(a,b)}= \frac{1}{\sqrt{2\pi^2 n}}\
\Lambda^{1,\alpha\beta}_{ab}\otimes P^{1+\alpha}Q^{\beta}\, ,
\end{equation}
where the matrix $\Lambda^{1,\alpha\beta}_{ab}$ has  components
$(\Lambda^{1,\alpha\beta}_{ab})_{cd}=
\lambda^{1,\alpha\beta}_{ab}\delta_{ac}\delta_{bd}$. This mode is
off-diagonal, unless $a=b$ and $\alpha=\beta=n$. It is considered to
be off-diagonal with respect to  the eigenvalue decomposition of $X$
if the $P,Q$ matrix component does not commute with $X$, or does not
$q$-commute with $Y$ and $Z$ according to the $F$-terms of the field
theory. Because in general the modes $\delta X$, $\delta Y$, $\delta
Z$ can mix, the problem of finding the correct modes can be
complicated. However, we can use the fact that this theory is
obtained from an orbifold of the ${\cal N}=4$ SYM theory to
diagonalize the problem. It turns out that after mixing is taken
into account, the Hamiltonian associated to these non-BPS modes is
\begin{equation}
\label{sbH} H^{1,\alpha\beta}_{ab} = \frac 12
|\Pi_{ab}^{1,\alpha\beta}|^2  + \frac{1}{2}
\left(\omega_{ab}^{1,\alpha\beta}\right)^2
|\lambda_{ab}^{1,\alpha\beta}|^2 \;,
\end{equation}
where the frequency of the mode turns out to be
\begin{equation}
\left(\omega_{ab}^{1,\alpha\beta}\right)^2 = 1 + \frac{g_{YM}^2}{2
\pi^2 n} |\vec x_a - \vec x_b^{-\alpha,\alpha-\beta}|
\;,
\end{equation}
In the end, we have to evaluate these energies using the
distribution of eigenvalues obtained in the previous section. The
$n^2$ non-BPS modes for two given points in the orbifold can be seen
as the $n^2$ possible straight lines connecting the first point with
the $n^2$ images of the second. This can be easily visualized in the
cone ${\mathbb C}^2/{\mathbb Z}_2$. In the Figure \ref{cono} we draw
the two possible straight lines connecting two points on the cone.

\myfig{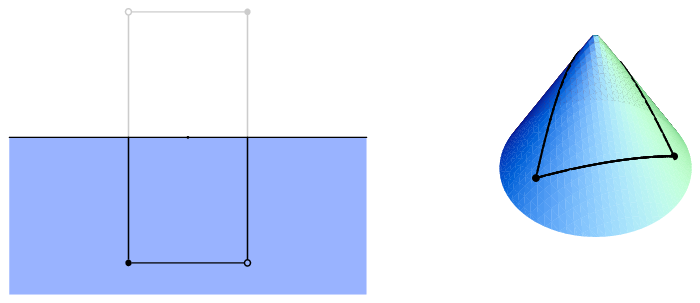}{10}{Straight lines between two points on
${\mathbb C}^2/{\mathbb Z}_2$.\label{cono}}

The interpretation of these modes as bits of massive strings goes
exactly as in the undeformed case \cite{Droplet}. According to the
AdS/CFT correspondence, in the limit of large 't Hooft coupling
$\lambda$, the system should be described in terms of classical
gravity with some string probes on the background. In this regime,
using normalized distances $r$, the energies of the non-BPS modes
are order
\begin{equation}
E_{ab}^{\alpha\beta}\sim\lambda^{1/2}|r_a -
r_b^{-\alpha,\alpha-\beta}|\,.
\end{equation}
where we have used the radius normalization found in the previous
section to rewrite the mass terms in terms of the 't Hooft coupling,
and unit vectors on the sphere $r_a$. At strong coupling, the
distance between the points dominates the energy of the string bits.
Finite small energy string bits (energy of order one) require very
small $|r_a - r_b^{-\alpha,\alpha-\beta}|$ and the segments between
$r_a$ and $r_b^{-\alpha,\alpha-\beta}$ become tangent to the
5-sphere. These segments have to be joined into closed loops because
we need to make a gauge invariant state.

We can consider these small segments as short bits of strings
localized on the geometrical $S^{5}/{\mathbb Z}_n\times{\mathbb
Z}_n$. In the case of strings on $AdS_5\times S^5$, it was shown in
\cite{BCV} that this description in terms of free short string bits
was useful for the BMN limit, but longer string bits should be
treated as interacting fields. Even though long string bits are
interacting, the leading order perturbative evaluation of their
energies gives the correct order of magnitude estimates for string
energies and one can also use them to get some intuition on the
eigenvalue geometry they are associated with.

\section{Energy for BMN states}
\label{spa}

In this section we repeat a similar analysis  to that of \cite{BCV}
to obtain an all-loop expression for the BMN operators energy in the
$q$-deformed theory. To do that, we compute the approximate energy
of the state resulting when we excite the ground state (\ref{ge})
with a BMN or BMN-like operator. BMN operators in marginal
deformations of ${\cal N}=4$ SYM have been already studied in the
literature \cite{Roiban,BC,FRT}, by considering appropriate
deformations of the usual spin chain describing the dilatation
operator. For instance, consider  the dilatation operator in the two
spin sector, at the 1-loop approximation, and with a deformation
parameter $q=r \exp(2\pi i\beta)$. For a real deformation $q=r$, the
dilatation operator turns out to be a periodic XXZ spin chain
\cite{Roiban,BC}. The anisotropy parameter is a function of the
deformation parameter such that $\Delta(r)\leq 1$ and then the spin
chain is always ferromagnetic. When the deformation parameter
presents a non-trivial phase, it is possible  to perform a position
dependent change of basis. In this particular new basis the
Hamiltonian of an XXZ spin chain is recovered and the phase of the
deformation is totally encoded in some twisted boundary conditions
and a deformation in the cyclicity condition \cite{BC}. Moreover, in
the case we are interested in, {\it i.e.} when the deformation
parameter is a pure phase
 $q=\exp(2\pi i\beta)$, the chain becomes an XXX spin chain with twisted boundary
 conditions. The twisting factor is $q^J$ where $J$ is the length of the spin
 chain.

It turns out that BMN-like operators with two different holomorphic
impurities are essentially identical to those of the undeformed
theory \cite{FRT} and the momentum numbers in the 1-loop energy are
shifted $k/J\to k/J \pm \beta$.  A consistent BMN limit for this
spin chain requires that $k \pm \beta J$ should be kept finite in
the $J\to\infty$ limit. In our case, it means that $k \pm J/n$
should be finite. So, we will consider the following BMN-like
operator for exciting the ground state (\ref{ge}),
\begin{equation}
O_{k}  \sim \sum_{l = 0}^J \exp( i \varphi l)\tr(Z^l X Z^{J-l} Y)\;,
\end{equation}
where $\varphi$ is the usual BMN phase $\varphi=  \frac {2\pi k}J$.
We take $Z$ to be of the form (\ref{qal3}) and the corresponding
eigenvalues $\lambda^3_a$ given by the distribution obtained in
section 3. On the other hand, fields $Y$ and $X$ include non-BPS
modes, that can be treated as creation operators in a quantum system
like (\ref{sbH}). We will refer to them as  $\lambda^{\! 1\
\alpha\beta \dagger}_{ab}$ and $\lambda^{\!2\ \alpha\beta \dagger
}_{ab}$. They satisfy the usual creation-annihilation algebra
\begin{equation}
[\lambda^{\alpha\beta }_{ab},\lambda^{\gamma\delta \dagger }_{cd}]
=\delta_{ad}\delta_{bc}\delta^{\alpha\gamma}\delta^{\beta\delta}\, .
\end{equation}
The above operator is represented by the following state
\begin{equation}
\ket{\psi_{k}} \sim \sum_{l = 0}^J e^{i\varphi l}
\sum_{\alpha,\beta}\sum_{a,b}
q^{l(\alpha-\beta+1)+(\alpha+1)(\beta-s)}(\lambda^3_a)^l
(\lambda^3_b)^{J-l} \lambda^{\! 2\ \alpha\beta
\dagger}_{ab}\lambda^{\! 1\ s-\alpha-1,s-\beta-1 \dagger}_{ba}
\hat\psi_0 \vac\;. \label{wf}
\end{equation}
The integer $s$ is the reduction modulo $n$ of the number of sites
$J$. We have written the wave function $\hat \psi_0$ in the
coordinate basis and the non-BPS modes as creation operators acting
on the vacuum $\vac$ of non-BPS modes.

We compute the energy of the above state, as an expectation value,
\begin{equation}
 E \sim \frac{\bra{\psi_{k}} H^{total} \ket{\psi_{k}}}{\braket{\psi_{k}}{\psi_{k}}}\;.
\end{equation}
The Hamiltonian (\ref{sbH}) tells us that the oscillators carry an
energy
\begin{equation}
E^{\alpha,\beta}_{ab}= \omega_{ab}^{1,\alpha\beta}
+\omega_{ba}^{2,r-\alpha-1,r-\beta-1} \;. \label{ode}
\end{equation}
Including the energy of the diagonal piece the total energy is
\begin{equation}
E^{total} = J + \langle E^{osc}\rangle\;.
\end{equation}
Then, we have to compute the average energy of the oscillators
(\ref{ode}) for the wave function (\ref{wf}). This results in the
following integral
\begin{equation}
\langle E^{osc}\rangle = \frac{\int \prod dx^c |\hat\psi_0|^2
\sum_{a,b,\alpha,\beta} |\sum_l e^{i l\varphi}
q^{l(\alpha-\beta+1)}(\lambda^3_a)^l (\lambda^3_b)^{J-l}|^2
 \left(\omega_{ab}^{1,\alpha\beta}
+\omega_{ba}^{2,r-\alpha-1,r-\beta-1}\right)} {\int \prod dx^c
|\hat\psi_0|^2 \sum_{a,b,\alpha,\beta} |\sum_l e^{i l\varphi}
q^{l(\alpha-\beta+1)}(\lambda^3_a)^l (\lambda^3_b)^{J-l}|^2}
\end{equation}
As it was done in the undeformed case \cite{BCV}, we can compute
this average by a saddle point approximation. The first thing to
notice is that both integral, in the denominator and in the
numerator are dominated by the configurations that maximize
$|\psi_0|^2$. This means that $\vec \lambda_a$ and $\vec
\lambda_{b}$ should be located exactly on the orbifolded sphere
$\tilde\Omega_5={S}^5/{\mathbb Z}_n\times {\mathbb Z}_n$. More
precisely, in the thermodynamic limit sums $\sum_a$ are converted
into integrals over the complex orbifold ${\mathbb C}^3/{\mathbb
Z}_n\times {\mathbb Z}_n$ which, using the distribution $\rho_0(x)$,
are reduced to integrals over $\tilde\Omega_5$.

\begin{equation}
\langle E^{osc}\rangle \simeq \frac{\int d\tilde\Omega_5
d{\tilde\Omega'}_5\sum_{\alpha,\beta} |\sum_l e^{i l\varphi}
q^{l(\alpha-\beta+1)}(\lambda^3)^l ({\lambda^3}')^{J-l}|^2
 \left(\omega_{(\lambda\lambda')}^{2,\alpha\beta}
+\omega_{(\lambda'\lambda)}^{1,r-\alpha-1,r-\beta-1}\right)} {\int
d\tilde\Omega_5 d{\tilde\Omega'}_5\sum_{\alpha,\beta} |\sum_l e^{i
l\varphi} q^{l(\alpha-\beta+1)}(\lambda^3)^l
({\lambda^3}')^{J-l}|^2}
\end{equation}
Using $\lambda^3=r_0\cos\theta\exp(i\phi)$,  the square of the sum
can be rephrased as
\begin{eqnarray}
 \label{suma}
\left|\sum_l e^{i l\varphi} q^{l(\alpha-\beta-1)}(\lambda^3)^l
({\lambda^3}')^{J-l} \right|^2 &=& r_0^{2J}\sum_{l}\sum_{l'}
(\cos\theta)^{l+l'}(\cos\theta')^{2J-l-l'}\times\\
&& ~ ~ ~ \exp\left[i(l-l')\left(\varphi + \phi
-\phi'+\frac{2\pi}{n}(\alpha-\beta+1)\right)\right] \nonumber
\end{eqnarray}

The BMN limit requires $J\to\infty$. This serves for improving the
saddle point approximation thanks to the extra powers of
$\cos\theta$ and $\cos\theta'$ in the angular integrals.  They are
maximized when  $|\lambda^3|$ and $|{\lambda^3}'|$ take their
maximum value on $\tilde\Omega_5$, {\it i.e.} when
$\cos\theta=\cos\theta'=1$. This corresponds to a geometrical
localization for the string with large angular momentum $J$ around a
null geodesic. Because of this localization the distances in the
frequencies of the non-BPS modes are simplified. For instance,
\begin{equation}
\label{fre} |\vec x - \vec {x'}^{-\alpha,\alpha-\beta}|^2 ~\to ~
|\lambda^3-q^{\beta-\alpha}{\lambda^3}'|^2 = 4 r_0^2
\sin^2\left(\frac 12 \left(\phi-\phi'+
{2\pi}(\alpha-\beta)/n\right)\right)\,.
\end{equation}
A similar simplification takes place for the other frequency,
leading to the same value (\ref{fre}). Moreover, in the large $J$
limit, we can approximate the sum over relative phases in
(\ref{suma}) by a delta function and compute easily the remaining
angular integrals. The result is that the amount $\phi-\phi'+
{2\pi}(\alpha-\beta)/n$ is sharply peaked at the value
$\varphi+2\pi/n$, and the energy of the oscillators is
\begin{equation}
\langle E^{osc}\rangle = 2\sqrt{1+\frac { 2 g^2_{YM}  r_0^2}{n
\pi^2} \sin^2\left( \frac{\varphi+2\pi/n}2\right)}\;,
\end{equation}
Using the value we obtained $r_0=\sqrt {\frac {mn^2}2}$ in the
section 3., $\varphi=  \frac {2\pi k}J$ and taking the large $J$
limit, the energy for the oscillators reduces to
\begin{equation}
\langle E^{osc}\rangle = 2\sqrt{1+g^2_{YM}  N  \left(\frac{
k}{J}+\frac 1n\right)^2}\;.
\end{equation}
which is exactly the energy of BMN string excitations, with the
appropriate normalization $g_{YM}^2=4\pi g_s$. From the gauge theory
side it corresponds  to an {\em all orders} result in perturbation
theory. This all loop gauge theory result has been already reported
in \cite{NP}. To obtain this result the authors of \cite{NP},
extended the arguments  used in \cite{SZ} for the undeformed case.
This includes using the equations of motion, but it is not clear if
contact terms could spoil the use of these relations. Also, string
states in the Lunin-Maldacena background have been considered by
many authors \cite{FRT,Fro,BRoi,deMel,Mateos}

We can also study in the same approximation BMN like states that are
not reduced to the BMN limit. For example, we can consider the
operator $\tr(Z^JY)$. These operators have also been considered in
\cite{Pen1,Pen2} (our results agree with those calculations).
It is easy to see that this trace is zero on
BPS states. In this case, we interpret $Y$ as the off-diagonal mode.
Doing the same analysis as above, we also localize on the circle
where $|z|=1$, and it turns out that the angle between the two
eigenvalues that $Y$ connects is exactly given by the argument of
$q$. This means that in the covering space, the mode $Y$ goes
between an eigenvalue of $\Lambda_Z$ and a particular one of it's
images. At first sight one might think that one is violating the
Gauss' law for the corresponding state, because it seems as if the
string state is an open string. However, this is a non-trivial loop
in the orbifold geometry, so the correct interpretation is that the
associated closed string state is actually in the twisted sector of
the orbifold. This is exactly as should be. Twisted states of an
abelian orbifold have non-trivial quantum symmetry quantum numbers.
These discrete quantum numbers of the states can be calculated
counting powers of $X,Y,Z$ modulo $n$ \cite{BL,BJL}. It is easy to
see that these jump assignments between images are all consistent
with the quantum symmetry quantum numbers of states in these papers.
Moreover, we can ask now under what conditions do some of these
states survive in the BMN limit. It's easy to show that if we take
the BMN limit along $Z$, we need the state to be built only of small
bits, and that forces us to have $\# Y-\# X=0 \mod(n)$. This
reproduces the BMN set of states of the orbifold $\BC^2/\BZ_n$,
exactly as one would have expected from spin chain model
considerations \cite{BC}.

\section{Discussion}
\label{dis} In this paper we studied BPS configurations in some
marginal deformations of SYM preserving ${\cal N}=1$ supersymmetry.
We focused our attention to the case in which the commutator in
the superpotential is deformed to a $q$-commutator, with  $q$ being
an $n$-th root of unity. In those cases, the gauge theory is
dual to the near horizon geometry of $N$ D-branes in the complex
orbifold ${\mathbb C}^3/{\mathbb Z}_n\times{\mathbb Z}_n$. The moduli
space is
described in terms of quantum mechanical
$q$-commuting matrices. In the large $N$ limit, this is reduced to
the study of density distributions of particles (interacting bosons)
in a 6-dimensional phase space, that results from studying the
moduli space of vacua of D-branes in the bulk. The ground state
corresponds to a spherical distribution of bosons in ${\mathbb
C}^3/{\mathbb Z}_n\times{\mathbb Z}_n$ and we identified this
distribution as the compact factor of the dual geometry. Moreover,
we considered non-BPS configurations by turning on non-trivial
$q$-commutators. We argued that, as in the undeformed pure ${\cal
N}=4$ SYM case, the energy of non-BPS modes is localized in this
density distribution identified with the compact factor of the
geometry, and they are considered as bits of massive strings. Thus,
one of the accomplishment of this paper is that we have an explicit
example of the  extension of qualitative picture \cite{Droplet} for
the appearance of spacetime geometry and locality in the strong  't
Hooft  coupling regime beyond the standard $AdS_5\times S^5$/${\cal
N}=4$ SYM.

Then, we proceeded as in \cite{BCV} to compute the BMN energies to
all orders in perturbation theory. The agreement with the string
theory computation is complete where it is expected. This
computation shows that the geometrical interpretation of the density
distribution of boson is not just a qualitative picture, but that it
can be used to reproduce concrete string theory calculations. We
want to emphasize that the approximations we made are  good only in
the strict BMN limit. If we were interested in near-BMN corrections,
we would have to deal with string bits of finite length and their
interactions. The energies of these string states are analytic
functions of the phase of $q$ (for $q$ unitary), so we can extend
them to non-rational values of $q$ and we get agreement with other
calculation of the corresponding energies.

Moreover, we have seen that the geometric picture given by this
distribution of eigenvalues accounts correctly for discrete quantum
numbers of string states, related to being in the twisted sector of
the orbifold. This is a necessary step to claim that we understand
the origin of geometry in a string theory setup that is not just
supergravity.

Obviously, since we have succeeded in showing that for certain
orbifolds of ${\cal N}=4 $ SYM theory we are able to find the
correct non-spherical horizon of the geometry, it is easy to
conjecture that it will work the same way for all such
supersymmetric orbifolds. Some recent progress in the better understanding of
the structure of the field theory at
orbifolds has been made recently in \cite{BeR,SadS}
It would be interesting to show that this
procedure works for other cases as well, such as the field theory
obtained by placing D3-branes at a conifold singularity \cite{KW}.
There is some preliminary evidence that this is possible in a large
class of examples \cite{BCor}.

\section*{Acknowledgements}
D. B. would like to thank R. Corrado,  J. Maldacena and S. Vazquez for various
discussions related to this problem.  D.H.C would like to thank C.
Herzog and R.Roiban for some useful conversations. D.B. work was
supported in part by a DOE OJI award, under grant DE-FG02-91ER40618.
D.B. would also like to thank the Institute for Advanced Study in Princeton and the Trinity College in Dublin for their hospitality while this work was being produced.
D.H.C. work was supported in part by NSF under grant No.
PHY99-07949, by Fundaci\'on Antorchas and by Fondecyt. Institutionalgrants to CECS of the Millennium Science Initiative, Fundaci\'on Andes, and the generous support by Empresas CMPC are also gratefully acknowledged.

 \end{document}